\documentclass[reviewcopy]{elsarticle}
\usepackage{natbib,longtable}
\usepackage{graphicx,array}
\usepackage[reviewcopy]{adndt}

\begin{document}

\begin{frontmatter}

  \title{Transition rates and radiative lifetimes of Ca I}

  \author[IOP]{Yanmei YU\corref{cor1}}
  \ead{ymyu@iphy.ac.cn}

  \author[UNR]{Andrei Derevianko}
 % \ead{andrei@unr.edu}

  \cortext[cor1]{Corresponding author.}
 % \fntext[X]{First author footnote.}
  %\fntext[Y]{Second author footnote.}

  \address[IOP]{Beijing National Laboratory for Condensed Matter Physics, Institute of Physics, Chinese Academy of Sciences, Beijing 100190,China}

  \address[UNR]{Department of Physics, University of Nevada, Reno, Nevada 89557, USA}

\begin{abstract}
We tabulate spontaneous emission rates for all possible 811 electric-dipole-allowed transitions between the  75 lowest-energy states of Ca I.
These involve the $4sns$ ($n=4-8$), $4snp$ ($n=4-7$), $4snd$ ($n=3-6$), $4snf$ ($n=4-6$), $3d^2$, $4p^2$, $3d4p$, and $4s5g$ electronic configurations.
We compile the transition rates by carrying out
 {\em ab initio} relativistic calculations
using the combined method of configuration interaction and many-body perturbation theory. The results are compared to the available literature values.
 The tabulated  rates can be useful in various applications, such as optimizing laser cooling in magneto-optical traps, estimating various systematic effects in optical clocks and evaluating static or dynamic polarizabilities and  long-range atom-atom interaction coefficients and related atomic properties.
\end{abstract}

%\begin{keyword}
%  \PACS 27.90.+b, 23.60.+e, 21.10.Tg, 21.30.Fe SHE; Mass formula;
%  $Q$-value; DDM3Y; WKB; VSS; Half life.
%\end{keyword}

\end{frontmatter}

\clearpage

\tableofcontents
\listofDtables
\listofDfigures

\clearpage

\section{Introduction}

Alkaline-earth atoms and divalent-like atoms (such as Yb and Hg) became a subject of interest to the cold atom community in the past decade. These  atoms possess two valence electrons outside a tightly bound core and in the LS coupling scheme,  the atomic states can be  classified by being either singlet or triplet states.  The availability of relatively wide spin-allowed and narrow spin-forbidden electric-dipole (E1) transitions enables stacking laser cooling
on both types of transitions, with the spin-allowed transitions used for the initial rapid cooling and spin-forbidden transitions - for reaching much lower Doppler-limit temperatures. Moreover, the
narrow inter-combination transitions, such as the $4s^2\,^1\!S_0-4s4p~^3\!P_1$ transition, can be used as an optical frequency reference~ \cite{Oates-OL-2000,Degenhardt-PRA-2005}.
The highly-forbidden $4s^2\,^1\!S_0-4s4p~^3\!P_0$ transition can be potentially used for an optical lattice clock scheme~\cite{DerKat11}.
However, due to peculiarities of its electronic structure, Ca, unlike other alkaline-earth atoms, has a relatively short lifetime in conventional magneto-optical traps (MOT). To improve the MOT efficiency, several re-pumping schemes were proposed and demonstrated~\cite{Mills-2017}. That work required reliable electric-dipole transition data for many transitions between the  75 lowest energy states of neutral Ca. Here, we compile the results of our computational work that served as a basis of the MOT performance analysis~\cite{Mills-2017}.
We anticipate that the tabulated data will be useful in multiple other contexts, such as estimating various systematic effects in optical clocks and computing  static or dynamic polarizabilities and  long-range inter-atomic interaction coefficients.

There have been a number of atomic-structure calculations for neutral calcium. The earlier work includes multi-configuration Hartree-Fock (MCHF) calculation \cite{Vaeck-JPB-1991} and semi-empirical model-potential calculations\cite{Mitroy-JPB-1993,Brage-PRA-1994,Laughlin-JPB-1996,Hansen-JPB-1999}. These computations provide  oscillator strengths for spin-allowed transitions for levels up to $4s10s$, $4s9p$ and $4s6d$. Most of them are non-relativistic with very limited numbers of low-lying levels treated with  {\em ab initio} relativistic methods. In particular, Fisher and Tachiev~\cite{Frischer-PRA-2003} reported energies and E1 transition rates for levels below $3d4p~^1F_3$. Porsev et al.~\cite{Porsev-PRA-2001} and Savukov and Johnson~\cite{Savukov-PRA-2002} computed the $4s^2-4s4p$, $4s4p-3d4s$, and $4s4p-4s5s$ transition rates  using a combination of configuration-interaction (CI) and many-body perturbation theory (MBPT) (referred to as the CI+MBPT method). The CI+MBPT method  results were in excellent agreement with  high-precision experimental values. This fact partially motivated our use of the relativistic  CI+MBPT method  for the present work.

%In this work, we tabulate electric-dipole transition rates for 811 transitions between the  75 lowest energy states of Ca I.  We also tabulate  radiative lifetimes of the above states. We use the relativistic CI+MBPT method. Based on our data, improved MOT schemes can been proposed. In addition, the Spontaneous emission rates data will aid in the determination of other important properties such as polarizabilities required for the evaluation of the blackbody radiation shift and seeking for the magic wavelength of atomic clock and long-range atomic interactions required in calculation of scattering lengths and interpretation of cold-collision data.

\section{Computational details}

%\subsection{Relativistic many-body calculations of atomic structure}
The CI+MBPT method employs a systematic formalism that combines advantages of both configuration interaction (CI) method and many-body perturbation theory (MBPT)~\cite{Dzuba-PRA-1996}. This method has been used extensively for evaluation of atomic properties (see, e.g., review~\cite{Derevianko-2011} for optical lattice clock applications and references therein). Relativistic effects are included exactly as the formalism starts from the Dirac equation and employs relativistic bi-spinor wave functions throughout the entire calculation.  In our treatment, the CI model space is limited to excitations of the two valence electrons. Contributions involving virtual excitations of core orbitals are treated within the MBPT. In this approach, we first solve for the valence electron orbitals and energies in the field of core electrons. The one-electron effective potential includes  both the frozen-cored Dirac-Hartree-Fock (DHF $V^{N-2}$) and self-energy (core-polarization) potentials. The self-energy correction is computed using second-order MBPT diagrams involving virtual core excitations. At the next step, the computed one-electron valence orbitals are used to diagonalize the atomic Hamiltonian in the model space of two valence electrons within the CI method.  The CI model-space Hamiltonian includes the residual (beyond DHF) Coulomb interaction between the valence electrons and also their core-polarization-mediated interaction. The latter was computed in the second-order MBPT. This step yields  two-electron wave-functions and energies. Finally, with the obtained wave-functions  we calculated the required electric-dipole matrix elements. In calculations of transition rates we used experimental energy intervals and the computed CI+MBPT matrix elements.

We used two independent CI+MBPT implementations: (i) by the Reno group (see the decription of the earlier version in Ref.~\cite{Derevianko-PRL-2011}) and (ii) the recently published package~\cite{Kozlov-CPC-2015}. The practical goal of the calculations was not reaching the best possible accuracy, but rather the generation of massive amounts of reliable data for the transition array involving 75 lowest-energy levels. The Reno code was run on a large basis set but without including core-polarization-mediated interaction in the CI Hamiltonian due to considerable computational costs. The production runs with package~\cite{Kozlov-CPC-2015} employed a smaller basis set (due to code limitations) but treated the correlation problem more fully.

While using the package~\cite{Kozlov-CPC-2015} we employed the one-electron basis set that included the $1s-17s$, $2p-17p$, $3d-17d$, $4f-17f$, and $5g-17g$ orbitals, where the core and $4s,\cdots,6f$ orbitals are DHF ones, while the remaining orbitals were represented by a B-spline basis set. The Reno code used the  dual-kinetic-balance basis set generated in the DHF $V^{N-2}$ potential using spherical cavity of 75 bohr radius \cite{Beloy-CPC-2008}.
The basis included orbitals with orbital angular momentum $\ell$ up to 6. The total number of positive-energy (in the Dirac sense) orbitals per partial wave was 40 with the 35 lowest-energy  orbitals used in the calculations.

For most states, the values of transition rates obtained with package~\cite{Kozlov-CPC-2015} were in closer agreement with the NIST recommended values due to more complete treatment of the correlation problem. The level of agreement was degraded  for states with the $4s6f$ electron configuration which was traced to our use of  small basis set due to   package~\cite{Kozlov-CPC-2015} limitations. For these $4s6f$ states, the values of transition rates obtained with the Reno code displayed a better agreement with the NIST data because of the larger basis set. In addition, due to the restriction on the number of eigenvalues in the package~\cite{Kozlov-CPC-2015}, we were not able to compute the states arising from the $4s8s$ configuration. Our final values combine the outputs of the two codes. While the bulk of the results comes from the package~\cite{Kozlov-CPC-2015}, rates for states  involving the $4s6f$ and $4s8s$ configurations are taken from the Reno code output in our tabulation.

The  spontaneous emission rate $A_{if}$ from the upper (initial) state $i$ to the lower (final) state $f$  was calculated as
\begin{equation}
%A_{if}=\frac{202.6684|\langle f||D||i\rangle|^2\times10^{8}}{\lambda^3 g_f},
%A_{if}= 2.02613 \times 10^{18}   \frac{|\langle f||D||i\rangle|^2}{(2J_f+1) \lambda_{fi}^3  } \mathrm{s}^{-1},
A_{if}= 2.02613 \times 10^{18}   \frac{|\langle i||D||f\rangle|^2}{(2J_i+1) \lambda^3  } \mathrm{s}^{-1},
\end{equation}
where $\langle i||D||f\rangle$ is the reduced matrix element of the electric-dipole operator in atomic units, the transition wavelength $\lambda$ is expressed in $\AA$, and $J_i$ is the total angular momentum of the upper (initial) state. The emission rate can be converted into the weighted oscillator strength using~\cite{Cowan}
\begin{equation}
gf=(2J_f+1)f_{fi}=-(2J_i+1)f_{if}=1.499\times10^{-16}A_{if}\lambda^2(2J_i+1),
\end{equation}
where $gf$ is weighted oscillator strength, $f_{if}$ is the emission oscillator strength (usually taken to be negative), $f_{fi}$ is the adsorption oscillator strength, and $J_f$ is the total angular momentum of the lower (final) state. We determined $\lambda$ from the NIST recommended energy values~\cite{NIST}. Finally, the total transition rate $A_\mathrm{total}$  from a given initial state is a sum of $A_{if}$ over all E1-allowed final states, with the resulting initial state lifetime $\tau= 1/A_\mathrm{total}$.

\section{Results and discussion }
We summarize the lifetime of the 75 states in Table 1 and compare with the available theoretical and experimental  values. Our data show better agreement with the experimental values, as compared with the other theoretical data. The computed spontaneous emission rates are compiled in Tables 2-6. For completeness, we list values obtained with  the electric dipole operator in both the length and velocity forms (gauges) in columns marked $L$ and $V$, respectively. Generally, the length and velocity gauges agree at a few percent level, except for some occasional large discrepancies for weak transitions. The length form values are in better overall agreement with the available literature data  and therefore we recommend using the length-form values.

We further assessed the quality of our calculation by comparing our values with other theoretical data \cite{Hansen-JPB-1999,Frischer-PRA-2003,Porsev-PRA-2001} and with the NIST recommended values~\cite{NIST} in Table 7. The other theoretical data are taken primarily from three sources. The first source is the non-relativistic semi-empirical calculation  by Hansen et al.~\cite{Hansen-JPB-1999}. They used the model potential and CI (MPCI) method to determine  rates for the LS allowed E1 transitions for levels up to $4s10s$. From a comparison with other semi-empirical work~\cite{Mitroy-JPB-1993,Brage-PRA-1994}, Hansen et al.~concluded that their data are more accurate because they have used larger basis size, larger cut-off radius, and more accurate model potential. In Table 7 we compare 80 transition rates with Hansen's results \cite{Hansen-JPB-1999}. For comparison, their weighted oscillator strength $gf$ values are converted to $A_{if}$ (in 1/s) as $A_{if}=6.67\times10^{15}gf/[(2J_i+1)\lambda^2]$, where $\lambda$ is in \AA.  The difference between our and the Hansen's results is better than 1\% for the $4s4p~^1P_1-4s^2~^1S_0$ and $4s4p~^3P_1-4s^2~^1S_0$ transitions and better than 50\% for majority of  other transitions. There are a few cases where the discrepancy between our and Hansen's data is worse than 50\% for some specific transitions decaying from highly excited states like $4sns$ with $n=6-8$, $4s7p$, $4s6d$, and $4s5d$. The second source is the {\em ab initio} relativistic MCHF calculations  by Fischer, et al.~\cite{Frischer-PRA-2003}. They report the transition rates between levels below $3d4p~^1F_3$, both allowed and spin forbidden. In Table 7, we compared 13 spin forbidden singlet-triplet transition rates with their results. Our values are in better agreement with the NIST data than the MCHF results, for example, for the $4s4p~^3P_1-4s^2~^1S_0$, $3d4p~^1D_2-3d4s~^3D_1$, $3d4p~^1D_2-3d4s~^3D_2$ transitions. The third source is the CI+MBPT calculation~\cite{Porsev-PRA-2001}. These authors reported rates for the spin-allowed $4s^2-4s4p$, $4s4p-3d4s$ transitions, and for the spin-forbidden $4s4p~^1P_1-4s5s~^3S_1$ and $4s4p~^3P_{1,2}-3d4s~^1D_2$ transitions. We find excellent agreement with their CI+MBPT results \cite{Porsev-PRA-2001}. Finally, we compared 99 transition rates with the NIST Atomic Spectra Database~\cite{NIST} in Table 7. In most cases, the discrepancy is better than 50\%. There are some instances when the discrepancy is worse than 50\%. In particular, for the $4s6f~^1F_0-3d4s~^1D_2$, $4s5p~^1P_1-4s^2~^1S_0$, and $4s4f~^1F_3-3d4s~^1D_2$ transitions, our data show good agreement with the MPCI theoretical data, but display large differences with the NIST recommended values.

%\subsection{Conclusion}
%We present theoretical transition probabilities for 811 lines and lifetimes corresponding to the first 75 excited states of Ca I. Our values for lifetimes are in reasonable agreement with experimental values. We have reported, for the first time, the relative values of transition probabilities of levels of $4sns$ ($n=6-8$),$4snp$ ($n=6-7$), $4snd$ ($n=4-6$), $4snf$ ($n=4-6$) and $4s5g$ configuration levels. These data are required for propose of new repumping scheme aiming higher cooling efficiency of MOT of Ca. These data are also useful for determination of polarizabilities and long-range interaction.

\ack We would like to thank Eric Hudson for motivating this work and Mikhail Kozlov for a valuable assistance with the  code~\cite{Kozlov-CPC-2015}.
Y.Y. thanks the University of Nevada, Reno for hospitality during her visit.
Y.Y. was supported by the National Natural Science Foundation of China under Grant No. 91536106, the NKRD Program of China (2016YFA0302104) and the CAS XDB21030300. A.D. was supported in part by the US National Science Foundation grant PHY-1607396.

\clearpage

\TableExplanation
%\section{Explanation of tables}

In Table I, we compile radiative E1  lifetimes of the  74 lowest-energy excited states of Ca I. In Tables 2-6, we tabulate our theoretical data for spontaneous emission rates of all possible 811 E1-allowed  transitions between the 75 lowest-energy states of Ca I. A comparison with other literature values is presented in Table 7.

\section*{Table 1. Radiative electric-dipole decay rates $A_\mathrm{total}$ and lifetimes $\tau$ of Ca I.}
\begin{tabular}{@{}p{2in}p{5in}@{}}
	First  column              & States. \\
	Second and third columns   & Energies in cm$^{-1}$, our CI+MBPT and the  NIST values. \\
	Fourth column              & Difference between our CI+MBPT and NIST energies  in \%. \\
	Fifth  column              & Total spontaneous emission rates $A_\mathrm{total}$  in $10^{8}$s$^{-1}$ for the state. \\
	Sixth and seventh column   & Lifetime of the state, our and the literature values. \\
	MPCI                       & Model Potential Configuration Interaction method.\\
	MCHF                       & Multi-Configuration Hartree-Fock method.\\
	Exp.                       & Experimental  values.\\
\end{tabular}

\section*{Table 2. Spontaneous emission rates of transitions originating from the $4sns$ ($n$=4-8)  states of Ca I.}
\begin{tabular}{@{}p{2in}p{5in}@{}}
First and Second columns   & Transition levels, the upper (initial) and the  lower (final) levels.\\
Third column               & Transition wavelength $\lambda$, in nm.\\
Fourth and Fifth columns   & Our values of spontaneous emission rates $A_{if}$ in $10^{8}\, \mathrm{s}^{-1}$ in  length ($L$) and velocity ($V$) forms of the E1 operator. \\
Sixth column               & Difference between the $L$ and $V$ column values defined as $(L-V)/L$ in \%. \\
Seventh column             & Branching ratio (BR) in \%, defined as $A_{if}/A_\mathrm{total}$.\\
\end{tabular}

\section*{Table 3. Spontaneous emission rates of transitions originating from the $4snp$ ($n$=4-7)  states of Ca I.}
\begin{tabular}{@{}p{2in}p{5in}@{}}
First and Second columns   & Transition levels, the upper (initial) and the  lower (final) levels.\\
Third column               & Transition wavelength $\lambda$, in nm.\\
Fourth and Fifth columns   & Our values of spontaneous emission rates $A_{if}$ in $10^{8}\, \mathrm{s}^{-1}$ in  length ($L$) and velocity ($V$) forms of the E1 operator. \\
Sixth column               & Difference between the $L$ and $V$ column values defined as $(L-V)/L$ in \%. \\
Seventh column             & Branching ratio (BR) in \%, defined as $A_{if}/A_\mathrm{total}$.\\
\end{tabular}

\section*{Table 4. Spontaneous emission rates of transitions originating from the $4snd$ ($n$=3-6) states of Ca I.}
\begin{tabular}{@{}p{2in}p{5in}@{}}
First and Second columns   & Transition levels, the upper (initial) and the  lower (final) levels.\\
Third column               & Transition wavelength $\lambda$, in nm.\\
Fourth and Fifth columns   & Our values of spontaneous emission rates $A_{if}$ in $10^{8}\, \mathrm{s}^{-1}$ in  length ($L$) and velocity ($V$) forms of the E1 operator. \\
Sixth column               & Difference between the $L$ and $V$ column values defined as $(L-V)/L$ in \%. \\
Seventh column             & Branching ratio (BR) in \%, defined as $A_{if}/A_\mathrm{total}$.\\
\end{tabular}

\section*{Table 5. Spontaneous emission rates of transitions originating from the $4snf$ ($n$=4-6) states of Ca I.}
\begin{tabular}{@{}p{2in}p{5in}@{}}
First and Second columns   & Transition levels, the upper (initial) and the  lower (final) levels.\\
Third column               & Transition wavelength $\lambda$, in nm.\\
Fourth and Fifth columns   & Our values of spontaneous emission rates $A_{if}$ in $10^{8}\, \mathrm{s}^{-1}$ in  length ($L$) and velocity ($V$) forms of the E1 operator. \\
Sixth column               & Difference between the $L$ and $V$ column values defined as $(L-V)/L$ in \%. \\
Seventh column             & Branching ratio (BR) in \%, defined as $A_{if}/A_\mathrm{total}$.\\
\end{tabular}

\section*{Table 6. Spontaneous emission rates of transitions originating from the $3d^2$, $4p^2$, $3d4p$, and $4s5g$ states of Ca I.}
\begin{tabular}{@{}p{2in}p{5in}@{}}
First and Second columns   & Transition levels, the upper (initial) and the  lower (final) levels.\\
Third column               & Transition wavelength $\lambda$, in nm.\\
Fourth and Fifth columns   & Our values of spontaneous emission rates $A_{if}$ in $10^{8}\, \mathrm{s}^{-1}$ in  length ($L$) and velocity ($V$) forms of the E1 operator. \\
Sixth column               & Difference between the $L$ and $V$ column values defined as $(L-V)/L$ in \%. \\
Seventh column             & Branching ratio (BR) in \%, defined as $A_{if}/A_\mathrm{total}$.\\
\end{tabular}

\section*{Table 7. Comparison of transition probabilities with  literature values.}
\begin{tabular}{@{}p{2in}p{5in}@{}}
First and Second columns   & Transition levels, the upper (initial) and the lower (final) levels.\\
Third column               & Spontaneous emission rates $A_{if}$ in units of $10^{8}\, \mathrm{s}^{-1}$ using the $L$-form of the E1 operator. \\
Forth and Fifth columns    & Other theoretical  $A_{if}$ values and the \% difference between this work and  other theoretical data, calculated as the difference of the third and forth column values divided by the third column value. \\
Sixth and Seventh columns  & NIST recommended $A_{if}$ value and the  \% difference between this work and NIST data, computed as the difference between the third and sixth column values divided by the third column value.\\
$L$                        & Length form of the electric-dipole operator. \\
MPCI                       & Model Potential Configuration Interaction method.\\
MCHF                       & Multi-Configuration Hartree-Fock method.\\
CI+MBPT                    & Configuration Interaction with Many-Body Perturbation Theory method.\\
NIST                       & NIST recommended values. \\
\end{tabular}

\clearpage
\LTright=0pt
\datatables % This command is necessary to get the table names in toc
\LTleft=0pt

\footnotesize

\begin{longtable}{@{\extracolsep\fill}ccccccc}
	\caption{Radiative electric-dipole decay rates $A_\mathrm{total}$ and lifetimes $\tau$ of Ca I.\label{tab:F}}\\[3pt]
	\hline \hline\noalign{\vskip3pt}
	State&\multicolumn{2}{c}{Energy (cm$^{-1}$)}&Diff.(\%)&$A_\mathrm{total}$($10^{8}$s$^{-1}$) &\multicolumn{2}{c}{$\tau$(ns)} \\[3pt]  \cline{2-3}\cline{6-7}
	&This work &NIST                     &        &                              &This work &Refs.      \\[3pt] \hline\noalign{\vskip3pt}
	\endfirsthead
	\multicolumn{7}{@{}c@{}}{Table  \ref{tab:F} contd\dots}\\  \hline \hline\noalign{\vskip3pt}
	State&\multicolumn{2}{c}{Energy (cm$^{-1}$)}&Diff.(\%)&$A_\mathrm{total}$($10^{8}$s$^{-1}$) &\multicolumn{2}{c}{$\tau$(ns)} \\[3pt]  \cline{2-3}\cline{6-7}
	&This work &NIST                    &        &                              &This work &Refs.      \\[3pt] \hline\noalign{\vskip3pt}
	\endhead
	\hline  \noalign{\vskip3pt}
	\multicolumn{6}{r}{\itshape Continued\dots}\\
	\endfoot
	\noalign{\vskip3pt}\hline
	\endlastfoot
	$4s4p~^3P_0$    &   15125.730  &   15157.900  &    0.2  &   0.0               &         $\infty$            &     \\
	$4s4p~^3P_1$    &   15179.500  &   15210.060  &    0.2  &   2.74 $\times10^{-5}$ &   3.65 $\times10^{5}$  &(3.4$\pm0.2$)$\times10^{5}$ Exp. \cite{Drozdowski-ZPD-1997}     \\
	                &              &              &         &                        &                        &4.25(10)$\times10^{5}$ Exp. \cite{Degenhardt-PRA-2005}   \\
	$4s4p~^3P_2$    &   15289.530  &   15315.940  &    0.2  &   0.0                &         $\infty$             &     \\
	$3d4s~^3D_1$    &   20822.630  &   20335.360  &    2.4  &   8.68 $\times10^{-3}$ &   1.15 $\times10^{3}$  &     \\
	$3d4s~^3D_2$    &   20840.440  &   20349.260  &    2.4  &   8.50 $\times10^{-3}$ &   1.18 $\times10^{3}$  &     \\
	$3d4s~^3D_3$    &   20868.160  &   20371.000  &    2.4  &   8.24 $\times10^{-3}$ &   1.21 $\times10^{3}$  &     \\
	$3d4s~^1D_2$    &   22449.920  &   21849.630  &    2.7  &   4.88 $\times10^{-6}$ & 2.05 $\times10^{6}$&2.3(5)$\times10^{6}$ Exp. \cite{Pasternack-JPB-1980}\\
\hline
\end{longtable}

\begin{longtable}{@{\extracolsep\fill}ccccccc}
 \caption{Spontaneous emission rates of transitions originating from the $4sns$ ($n$=4-8)  states of Ca I.\label{tab:A}}\\ %[3pt]
 \hline \hline\noalign{\vskip3pt}
  \multicolumn{2}{c}{Transition}&$\lambda$(nm) &\multicolumn{2}{c}{$A_{if}$($10^{8}$s$^{-1}$)}&Diff. (\%) &BR.(\%)  \\[3pt]  \cline{1-2}\cline{4-5}
  Initial &Final                &              &$L$&$V$&                                    &         \\[3pt] \hline\noalign{\vskip3pt}
\endfirsthead
 \multicolumn{7}{@{}c@{}}{Table  \ref{tab:A} contd\dots}\\  \hline \hline\noalign{\vskip3pt}
  \multicolumn{2}{c}{Transition}&$\lambda$(nm) &\multicolumn{2}{c}{$A_{if}$($10^{8}$s$^{-1}$)}&Diff. (\%) &BR.(\%)  \\[3pt]  \cline{1-2}\cline{4-5}
  Initial &Final                &              &$L$&$V$&                                    &         \\[3pt]   \hline\noalign{\vskip3pt}
\endhead
 \hline  \noalign{\vskip3pt}
 \multicolumn{7}{r}{\itshape Continued\dots}\\
\endfoot
 \noalign{\vskip3pt}\hline
\endlastfoot
$4s5s~^3S_1$   & $4s4p~^3P_0$   &    610.441 &  9.246E-02 &  9.001E-02 &     2.65 &  11.23 \\
$4s5s~^3S_1$   & $4s4p~^3P_1$   &    612.391 &  2.759E-01 &  2.685E-01 &     2.67 &  33.51 \\
$4s5s~^3S_1$   & $4s4p~^3P_2$   &    616.388 &  4.549E-01 &  4.426E-01 &     2.71 &  55.26 \\
$4s5s~^3S_1$   & $4s4p~^1P_1$   &   1267.877 &  5.911E-06 &  6.068E-06 &    -2.66 &   0.00 \\[1ex]
$4s5s~^1S_0$   & $4s4p~^3P_1$   &    552.267 &  1.694E-05 &  1.826E-05 &    -7.78 &   0.01 \\
$4s5s~^1S_0$   & $4s4p~^1P_1$   &   1034.665 &  2.293E-01 &  2.217E-01 &     3.34 &  99.99 \\[1ex]
$4s6s~^3S_1$   & $3d4p~^3F_2$   &   2108.019 &  2.378E-05 &  2.360E-05 &     0.77 &   0.01 \\
$4s6s~^3S_1$   & $3d4p~^1D_2$   &   2155.716 &  1.673E-04 &  1.664E-04 &     0.53 &   0.05 \\
\end{longtable}

\LTright=0pt
\LTleft=0pt
\begin{longtable}{@{\extracolsep\fill}ccccccc}
  \caption{Spontaneous emission rates of transitions originating from the $4snp$ ($n$=4-7)  states of Ca I.\label{tab:B}}\\ %[3pt]
 \hline \hline\noalign{\vskip3pt}
  \multicolumn{2}{c}{Transition}&$\lambda$(nm) &\multicolumn{2}{c}{$A_{if}$($10^{8}$s$^{-1}$)}&Diff. (\%) &BR.(\%)  \\[3pt]  \cline{1-2}\cline{4-5}
  Initial &Final                &              &$L$&$V$&                                    &         \\[3pt] \hline\noalign{\vskip3pt}
\endfirsthead
 \multicolumn{7}{@{}c@{}}{Table  \ref{tab:B} contd\dots}\\  \hline \hline\noalign{\vskip3pt}
  \multicolumn{2}{c}{Transition}&$\lambda$(nm) &\multicolumn{2}{c}{$A_{if}$($10^{8}$s$^{-1}$)}&Diff. (\%) &BR.(\%)  \\[3pt]  \cline{1-2}\cline{4-5}
  Initial &Final                &              &$L$&$V$&                                    &         \\[3pt]   \hline\noalign{\vskip3pt}
\endhead
 \hline  \noalign{\vskip3pt}
 \multicolumn{7}{r}{\itshape Continued\dots}\\
\endfoot
 \noalign{\vskip3pt}\hline
\endlastfoot
$4s4p~^3P_1$   & $4s^2~^1S_0$   &    657.460 &  2.738E-05 &  3.531E-05 &   -28.98 & 100.00 \\[1ex]
$4s4p~^1P_1$   & $4s^2~^1S_0$   &    422.792 &  2.170E+00 &  2.327E+00 &    -7.20 & 100.00 \\
$4s4p~^1P_1$   & $3d4s~^3D_1$   &   3014.827 &  3.767E-08 &  4.832E-08 &   -28.27 &   0.00 \\
$4s4p~^1P_1$   & $3d4s~^3D_2$   &   3027.514 &  1.000E-07 &  3.825E-07 &  -282.52 &   0.00 \\
$4s4p~^1P_1$   & $3d4s~^1D_2$   &   5547.327 &  5.341E-05 &  2.287E-06 &    95.72 &   0.00 \\[1ex]
$4s5p~^3P_0$   & $4s5s~^3S_1$   &   1996.729 &  7.103E-02 &  7.143E-02 &    -0.55 &  40.46 \\
$4s5p~^3P_0$   & $3d4s~^3D_1$   &    616.814 &  1.045E-01 &  1.102E-01 &    -5.38 &  59.54 \\[1ex]
$4s5p~^3P_1$   & $4s5s~^3S_1$   &   1993.919 &  7.055E-02 &  7.096E-02 &    -0.58 &  40.15 \\
\end{longtable}

\LTright=0pt
\LTleft=0pt
\begin{longtable}{@{\extracolsep\fill}ccccccc}
 \caption{Spontaneous emission rates of transitions originating from the $4snd$ ($n$=3-6) states of Ca I.\label{tab:C}}\\ %[3pt]
 \hline \hline\noalign{\vskip3pt}
  \multicolumn{2}{c}{Transition}&$\lambda$(nm) &\multicolumn{2}{c}{$A_{if}$($10^{8}$s$^{-1}$)}&Diff. (\%) &BR.(\%)  \\[3pt]  \cline{1-2}\cline{4-5}
  Initial &Final                &              &$L$&$V$&                                    &         \\[3pt] \hline\noalign{\vskip3pt}
\endfirsthead
 \multicolumn{7}{@{}c@{}}{Table  \ref{tab:C} contd\dots}\\  \hline \hline\noalign{\vskip3pt}
  \multicolumn{2}{c}{Transition}&$\lambda$(nm) &\multicolumn{2}{c}{$A_{if}$($10^{8}$s$^{-1}$)}&Diff. (\%) &BR.(\%)  \\[3pt]  \cline{1-2}\cline{4-5}
  Initial &Final                &              &$L$&$V$&                                    &         \\[3pt]   \hline\noalign{\vskip3pt}
\endhead
 \hline  \noalign{\vskip3pt}
 \multicolumn{7}{r}{\itshape Continued\dots}\\
\endfoot
 \noalign{\vskip3pt}\hline
\endlastfoot
$3d4s~^3D_1$   & $4s4p~^3P_0$   &   1931.449 &  4.891E-03 &  4.940E-03 &    -0.99 &  56.36 \\
$3d4s~^3D_1$   & $4s4p~^3P_1$   &   1951.105 &  3.564E-03 &  3.604E-03 &    -1.11 &  41.06 \\
$3d4s~^3D_1$   & $4s4p~^3P_2$   &   1992.262 &  2.239E-04 &  2.271E-04 &    -1.46 &   2.58 \\[1ex]
$3d4s~^3D_2$   & $4s4p~^3P_1$   &   1945.828 &  6.467E-03 &  6.537E-03 &    -1.07 &  76.09 \\
$3d4s~^3D_2$   & $4s4p~^3P_2$   &   1986.760 &  2.032E-03 &  2.060E-03 &    -1.37 &  23.91 \\[1ex]
$3d4s~^3D_3$   & $4s4p~^3P_2$   &   1978.216 &  8.240E-03 &  8.340E-03 &    -1.22 & 100.00 \\[1ex]
$3d4s~^1D_2$   & $4s4p~^3P_1$   &   1506.122 &  4.059E-06 &  3.854E-06 &     5.05 &  83.12 \\
$3d4s~^1D_2$   & $4s4p~^3P_2$   &   1530.529 &  8.245E-07 &  4.298E-07 &    47.87 &  16.88 \\
\end{longtable}

\LTright=0pt
\LTleft=0pt

\begin{longtable}{@{\extracolsep\fill}ccccccc}
 \caption{Spontaneous emission rates of transitions originating from the $4snf$ ($n$=4-6) states of Ca I.\label{tab:D}}\\ %[3pt]
 \hline \hline\noalign{\vskip3pt}
  \multicolumn{2}{c}{Transition}&$\lambda$(nm) &\multicolumn{2}{c}{$A_{if}$($10^{8}$s$^{-1}$)}&Diff. (\%) &BR.(\%)  \\[3pt]  \cline{1-2}\cline{4-5}
  Initial &Final                &              &$L$&$V$&                                    &         \\[3pt] \hline\noalign{\vskip3pt}
\endfirsthead
 \multicolumn{7}{@{}c@{}}{Table  \ref{tab:D} contd\dots}\\  \hline \hline\noalign{\vskip3pt}
  \multicolumn{2}{c}{Transition}&$\lambda$(nm) &\multicolumn{2}{c}{$A_{if}$($10^{8}$s$^{-1}$)}&Diff. (\%) &BR.(\%)  \\[3pt]  \cline{1-2}\cline{4-5}
  Initial &Final                &              &$L$&$V$&                                    &         \\[3pt]   \hline\noalign{\vskip3pt}
\endhead
 \hline  \noalign{\vskip3pt}
 \multicolumn{7}{r}{\itshape Continued\dots}\\
\endfoot
 \noalign{\vskip3pt}\hline
\endlastfoot
$4s4f~^3F_2$   & $4s5s~^3S_1$   &    940.671 &  2.802E-13 &  1.357E-13 &    51.58 &   0.00 \\
$4s4f~^3F_2$   & $4s4d~^1D_2$   &   2052.579 &  1.385E-06 &  1.352E-06 &     2.37 &   0.00 \\
$4s4f~^3F_2$   & $4s4d~^3D_1$   &   2261.415 &  8.393E-02 &  8.413E-02 &    -0.23 &  32.22 \\
$4s4f~^3F_2$   & $4s4d~^3D_2$   &   2263.293 &  1.559E-02 &  1.562E-02 &    -0.21 &   5.98 \\
$4s4f~^3F_2$   & $4s4d~^3D_3$   &   2266.155 &  4.422E-04 &  4.429E-04 &    -0.18 &   0.17 \\
$4s4f~^3F_2$   & $4p^2~^3P_1$   &   2698.764 &  2.194E-06 &  2.208E-06 &    -0.60 &   0.00 \\
$4s4f~^3F_2$   & $4p^2~^3P_2$   &   2763.462 &  6.099E-07 &  6.172E-07 &    -1.21 &   0.00 \\
$4s4f~^3F_2$   & $4s6s~^3S_1$   &   5896.331 &  2.241E-13 &  2.165E-13 &     3.39 &   0.00 \\
\end{longtable}

\LTright=0pt
\LTleft=0pt
\begin{longtable}{@{\extracolsep\fill}ccccccc}
 \caption{Spontaneous emission rates of transitions originating from the $3d^2$, $4p^2$, $3d4p$, and $4s5g$ states of Ca I.\label{tab:E}}\\ %[3pt]
 \hline \hline\noalign{\vskip3pt}
  \multicolumn{2}{c}{Transition}&$\lambda$(nm) &\multicolumn{2}{c}{$A_{if}$($10^{8}$s$^{-1}$)}&Diff. (\%) &BR.(\%)  \\[3pt]  \cline{1-2}\cline{4-5}
  Initial &Final                &              &$L$&$V$                                     &                                   &         \\[3pt] \hline\noalign{\vskip3pt}
\endfirsthead
 \multicolumn{7}{@{}c@{}}{Table  \ref{tab:E} contd\dots}\\  \hline \hline\noalign{\vskip3pt}
  \multicolumn{2}{c}{Transition}&$\lambda$(nm) &\multicolumn{2}{c}{$A_{if}$($10^{8}$s$^{-1}$)}&Diff. (\%) &BR.(\%)  \\[3pt]  \cline{1-2}\cline{4-5}
  Initial &Final                &              &$L$&$V$&                                    &         \\[3pt]   \hline\noalign{\vskip3pt}
\endhead
 \hline  \noalign{\vskip3pt}
 \multicolumn{7}{r}{\itshape Continued\dots}\\
\endfoot
 \noalign{\vskip3pt}\hline
\endlastfoot
$3d4p~^3F_2$   & $4s5s~^3S_1$   &   2386.094 &  1.489E-05 &  1.490E-05 &    -0.06 &   0.00 \\
$3d4p~^3F_2$   & $3d4s~^3D_1$   &    649.558 &  2.250E-01 &  2.062E-01 &     8.37 &  56.13 \\
$3d4p~^3F_2$   & $3d4s~^3D_2$   &    650.145 &  4.840E-02 &  4.514E-02 &     6.74 &  12.07 \\
$3d4p~^3F_2$   & $3d4s~^3D_3$   &    651.065 &  7.591E-04 &  6.482E-04 &    14.60 &   0.19 \\
$3d4p~^3F_2$   & $3d4s~^1D_2$   &    720.419 &  1.267E-01 &  1.435E-01 &   -13.25 &  31.61 \\[1ex]
$3d4p~^3F_3$   & $3d4s~^3D_2$   &    646.435 &  3.367E-01 &  3.086E-01 &     8.36 &  88.32 \\
$3d4p~^3F_3$   & $3d4s~^3D_3$   &    647.345 &  4.447E-02 &  4.113E-02 &     7.51 &  11.67 \\
$3d4p~^3F_3$   & $3d4s~^1D_2$   &    715.867 &  5.039E-05 &  4.493E-05 &    10.83 &   0.01 \\
\end{longtable}

\LTright=0pt
\LTleft=0pt

\begin{longtable}{@{\extracolsep\fill}ccccccc}
 \caption{Comparison of transition probabilities with  literature values. \label{tab:G}}\\ %[3pt]
 \hline \hline\noalign{\vskip3pt}
  Final state&Initial state     &This work ($L$, $10^{8}$s$^{-1}$) & Other theory ($10^{8}$s$^{-1}$)&Diff.&NIST &Diff. (\%) \\[3pt] \hline\noalign{\vskip3pt}
\endfirsthead
 \multicolumn{7}{@{}c@{}}{Table  \ref{tab:G} contd\dots}\\  \hline \hline\noalign{\vskip3pt}
  Final state&Initial state     &This work ($L$, $10^{8}$s$^{-1}$) & Other theory ($10^{8}$s$^{-1}$)&Diff.&NIST &Diff. (\%) \\[3pt]   \hline\noalign{\vskip3pt}
\endhead
 \hline  \noalign{\vskip3pt}
 \multicolumn{7}{r}{\itshape Continued\dots}\\
\endfoot
 \noalign{\vskip3pt}\hline
\endlastfoot
$4s4p~^1P_1$   & $4s^2~^1S_0$   &  2.170E+00 &  2.171E+00 MPCI \cite{Hansen-JPB-1999}&  -0.02 &  2.18E+00&  -0.44 \\
               &                &            &  2.15E+00 CI+MBPT \cite{Porsev-PRA-2001}& 0.9  &           &        \\
$4s4p~^3P_1$   & $4s^2~^1S_0$   &  2.738E-05 &  2.75E-05 CI+MBPT \cite{Porsev-PRA-2001}&0.43   &  2.6E-05&   5.30 \\
               &                &            &  2.098E-05 MCHF \cite{Frischer-PRA-2003}&23.37  &           &       \\
$4s4p~^1P_1$   & $3d4s~^1D_2$   &  5.341E-05 &  5.274E-05 MPCI \cite{Hansen-JPB-1999}& 1.26 &           &        \\
               &                &            &  5.70E-05 CI+MBPT \cite{Porsev-PRA-2001}&6.72  &           &        \\
$4s6f~^1F_3$   & $3d4s~^1D_2$   &  8.390E-02 &  8.570E-02 MPCI \cite{Hansen-JPB-1999} &  -2.10 &  9.0E-01& -90.68 \\
$4s5p~^1P_1$   & $4s^2~^1S_0$   &  7.505E-03 &  5.850E-03 MPCI \cite{Hansen-JPB-1999}&  28.30 &  2.7E-03& 177.96 \\
\end{longtable}

%\end{thebibliography}
\end{document}